\documentclass[twocolumn,showpacs,preprintnumbers,amsmath,amssymb,prl]{revtex4}

\usepackage{graphicx}% Include figure files
\usepackage{dcolumn}% Align table columns on decimal point
\usepackage{bm}% bold math

\newcommand{\be}{\begin{equation}}
\newcommand{\ee}{\end{equation}}
\newcommand{\bea}{\begin{eqnarray}}
\newcommand{\eea}{\end{eqnarray}}
\newcommand{\df}{{\rm d}}

\begin{document}

\preprint{Classical capacity of the noisy \ldots}

\title{Classical capacity of the noisy bosonic channel \\
and the bosonic minimum output entropy conjecture}

\author{Antonio Mecozzi}%
\email{antonio.mecozzi@univaq.it}
\affiliation{%
Department of Physical and Chemical Sciences, University of L'Aquila, via Vetoio 1, 67100 L'Aquila, Italy}

\date{\today}

\begin{abstract}
We consider a line with noise in the simplest case. Loss does not
add noise \cite{Giovannetti}. Amplification via phase insensitive
amplifiers do add noise. A lower bound of this capacity is the quantum analog to the Shannon capacity of a linear channel with additive white Gaussian noise, namely the difference of the Von Neumann entropy of the signal plus noise
at the output of the line and the entropy of the noise alone. We show that this expression is indeed the capacity for the case of an amplifier with infinitesimal gain $G = 1+\epsilon$, and for a cascade of an amplifier with arbitrary gain and a large loss, such that the overall gain of the cascade is infinitesimal.

\end{abstract}

%\keywords{}
\maketitle

There are few channels whose quantum capacity is known. One is the
lossy channel, whose capacity has been shown in ref.
\onlinecite{Giovannetti} to be equal to the capacity of a lossless
channel with the same average number of photons.  The result of
the lossy channel shows that the loss does not introduce any
effective noise to the transmission of information, in the sense
that the capacity is only limited by the average number of
received photons, no matter what the average number of photons
before transmission was.  The next step is the case the quantum equivalent of a noisy channel.  The
simplest noisy channel is classical information theory is a
channel with additive white Gaussian noise (AWGN) \cite{Shannon}.

One of the most known and celebrated Shannon result is the
expression for the capacity of such channel with a constraint of
an assigned average energy of the signal. Shannon showed that the capacity is attained for a Gaussian distribution of levels, and is the difference of the entropy of the received signal plus noise minus the entropy of the
noise alone. This result is valid only for a specific channel,
namely a linear channel with additive white Gaussian noise.  The quantum analogue of a channel with AWGN is a channel with a linear phase insensitive amplifier. Like the classical counterpart, the capacity
of such a channel has be conjectured to be again the difference of the Von Neumann entropy
of the received signal corrupted by the amplified spontaneous
emission (ASE) noise of the amplifier when the input state is Gaussian distributed, minus the Von Neumann entropy
of the ASE noise alone. This conjecture has been proven if the class of input states is constrained to classical input states and Gaussian states \cite{Holevo}.

In this note, we give an alternative proof of the conjecture for the known case of classical and Gaussian input states. In addition, we \textit{directly} prove the conjecture, without constraining the class of input states, for the cases of an amplifier with infinitesimal gain, and for a cascade of an amplifier with arbitrary gain and a large loss, such that the overall gain of the cascade is infinitesimal. This means that the local minima of the rate of entropy production for an arbitrary input state discussed in \cite{Lloyd} are in the case of infinitesimal gain either all zeros of the Lagrange equations or they have larger rate of production of entropy of the zeros.

\section{Holevo information of an ideal lossless channel}

Let us begin our analysis assuming that coherent states are
transmitted. The Holevo information is
\be \chi = S(\rho) - \int \frac{\df^2 \alpha}{\pi} p(\alpha)
S(|\alpha\rangle \langle \alpha |) = S(\rho), \label{10} \ee
where
\be S(\rho) = - \langle \rho \log(\rho) \rangle =
\mathrm{Tr}\left[ \rho \log(\rho) \right]. \label{20} \ee
Let us assume that \cite{Giovannetti}
\be p(\alpha) = \frac{\exp\left(-|\alpha|^2/N\right)}{N},
\label{30} \ee
such that
\be \rho = \int \frac{\df^2 \alpha}{\pi}
\frac{\exp\left(-|\alpha|^2/N\right)}{N} |\alpha \rangle \langle
\alpha |. \label{40} \ee
The density matrix (\ref{40}) is that of the ASE from a phase
insensitive amplifier \cite{Mecozzi} and vacuum state input.  The
entropy (\ref{20}) may be evaluated using the convenient
decomposition of $\rho$,
\be \rho = \sum_m \sum_n | m \rangle \langle m| \rho |n \rangle
\langle n|. \label{60} \ee
Using
\be \langle \alpha | n \rangle =
\exp\left(-\frac{|\alpha|^2}{2}\right)
\frac{\alpha^{*n}}{\sqrt{n!}}, \label{70} \ee
into (\ref{40}) and defining $\alpha = R e^{i \phi}$ we have
\bea \langle m| \rho |n \rangle &=& \int
\frac{\exp\left[-(N+1)R^2/N\right]}{N} \nonumber
\\
&& \times R^{m+n} e^{i(n-m) \phi} \frac{2 R \df R \df \phi}{2
\pi}. \label{80} \eea
Performing the integral, we obtain
\be \langle m | \rho | n \rangle = \frac {N^n}{(1+N)^{n+1}}
\delta_{m,n}, \label{90} \ee
hence (\ref{60}) becomes
\be \rho = \sum_n \frac {N^n}{(1+N)^{n+1}} | n\rangle \langle n|.
\label{100} \ee
The density matrix (\ref{40}) is diagonal in both coherent state
and number state basis. Using the above expression, it is easy to
show that the von Neumann entropy is
\be S(\rho) = g(N),  \label{110} \ee
where
\bea g(x) &=& (x+1) \log(x+1) - x \log x \nonumber \\
&=& \log(1+x) + x \log\left(1 + 1/x\right), \label{150} \eea
this being a well known result dating back to the work of James P.
Gordon in the early sixties \cite{Gordon62,Gordon64}. The second term in (\ref{10}) is
obviously zero because proportional to the entropy of a pure
state, hence we have the final result \cite{Giovannetti}
\be \chi = g(N). \label{120} \ee

\section{Holevo information of a channel with a phase insensitive amplifier}

Assume now that we want to transmit on an amplified channel.  An ideal phase sensitive amplifier can be described by a unitary operator, so that the capacity of such a channel is the same as the ideal lossless channel, Eq. (\ref{120}). Let us therefore concentrate on phase insensitive amplifiers. The density matrix at output of a phase insensitive amplifier with
gain $G$ (or with average ASE photons $G-1$) with the coherent
state input $|\alpha\rangle$ is
\bea \rho(\alpha) &=& \int \df^2 \alpha' P_{\mathrm{amp},G} (\alpha'-\sqrt{G} \alpha ) |\alpha' \rangle \langle \alpha'|, \\
P_{\mathrm{amp},G} \left(\alpha\right) &=& \frac{\exp\left[-|\alpha|^2/(G-1)\right]}{\pi(G-1)}. \label{50} \eea
Defining the displacement operator \cite{Glauber}
\bea D(\alpha) &=& \exp\left(\alpha a^\dagger- \alpha^* a\right) \nonumber \\
&=& \exp\left(-|\alpha|^2/2\right) \exp\left(\alpha a^\dagger\right) \exp\left(- \alpha^* a\right), \eea
we obtain
\be \rho(\alpha) = D(\sqrt G \alpha) \rho_\mathrm{ASE}(G) D^\dagger(\sqrt G \alpha), \label{55} \ee
where
\be \rho_\mathrm{ASE}(G) = \int \df^2 \alpha' P_{\mathrm{amp},G} (\alpha') |\alpha' \rangle \langle \alpha'|, \ee
is the density matrix corresponding to ASE alone, i.e. to a vacuum state at input. Notice that $D^\dagger(\alpha)$ is a unitary operator (a rotation in Hilbert space) because $D^\dagger(\alpha) D(\alpha) = 1$.

If the input is ASE-like as given by Eq. (\ref{40}) we have
\bea \rho_\mathrm{out} &=& \int \frac{\df^2 \alpha}{\pi}
\frac{\exp\left(-|\alpha|^2/N\right)}{N} \rho(\alpha) \nonumber \\
&=& \int \frac{\df^2 \alpha}{\pi} \frac{\exp\left[-
|\alpha|^2/(G N+G-1)\right]}{GN+G-1} |\alpha \rangle \langle
\alpha | \nonumber \\
&=& \rho_\mathrm{ASE}(NG+G). \label{130} \eea
A comparison with (\ref{40}) and (\ref{110}) shows that the entropy of the amplified state is
\be S(\rho_\mathrm{out}) = g(NG+G-1), \label{140} \ee
which tends to $g(x) \simeq \log(1+x) +1$ for $x \gg 1$. We need
now to compute the entropy of the state at the amplifier output
when the input is a coherent state $\alpha$, described by the
density matrix $\rho(\alpha)$. The state is described by
(\ref{50}). For every $\alpha$, the state distribution is obtained by that corresponding to a vacuum state at input by the application of the unitary operator $D(\sqrt G \alpha)$. Since the application of a unitary operator to a quantum state does not change the entropy,
we have
\be S[\rho(\alpha)] = g(G-1). \label{160} \ee
The Holevo information $\chi$ can be written immediately in this
case,
\be \chi(G) = g(NG+G-1) - g(G-1). \label{170} \ee
If we define the average output signal photon number $
N_\mathrm{out} = N G$ and the average output ASE photon number
$N_\mathrm{ASE}= G-1$ we may write
\bea \chi = g(N_\mathrm{out}+N_\mathrm{ASE}) - g(N_\mathrm{ASE}).
\label{180} \eea
Expression (\ref{180}) was conjectured by J. P. Gordon in 1964 \cite{Gordon64} and is more general than (\ref{170}), as it may be easily shown to be as well valid when there is loss together
with gain in the line connecting the transmitter to the receiver.
The results (\ref{170}) and (\ref{180}) is similar to the Shannon
result for the capacity of a classic channel corrupted by AWGN \cite{Shannon}:
the channel capacity is equal to the entropy of the signal
corrupted by noise less the entropy of the noise alone, that is
the entropy of the line output corresponding to zero signal, i.e.
a vacuum state, input. This suggests that, in this case, the Holevo information may be also the capacity.

Let us prove this statement for (\ref{170}) for simplicity, but the result can be
easily shown in the most general case of (\ref{180}). So, the first
term maximizes the entropy because for a fixed average input
(hence output) photon number, the output Bose-Einstein
distribution maximizes the entropy, for any choice of input
states. On the other hand, the second term, the conditional
entropy, is a minimum, if a coherent state input minimizes the output entropy of a phase insensitive amplifier, for any choice of input states. If this second statement is true in general, then the information rate (\ref{170}) is also the capacity.

\section{Classical and Gaussian input states}

Let concentrate the analysis on Eq. (\ref{170}) for simplicity, but the results can be
easily shown in the most general case of (\ref{180}). If the transmitted states have a positive Glauber-Sudarshan $P$-representation
\be \rho_0 = \int \df^2 \alpha P(\alpha) |\alpha\rangle \langle \alpha|, \ee
the output is
\be \rho = \int \df^2 \alpha P(\alpha) \rho(\alpha), \ee
and the entropy is
\be S(\rho) \ge \int \df^2 \alpha P(\alpha) S\left[\rho(\alpha) \right] = g(G-1), \label{210} \ee
where we used the concavity of the entropy \cite{Wehrl}, holding because of the positive-definiteness of $P(\alpha)$ insured by the classical nature of the state, together with (\ref{50}) and (\ref{160}). So, if we restrict the input states to classical states, (\ref{170}) is the capacity.

The class of states can be extended to all non-classical input states that give an amplifier output that can be obtained by applying a unitary transformation to the output of a phase independent amplifier with gain $G' \ge G$ and a classical state input. This is because a) for classical states inputs the minimum entropy output is given by the vacuum state, b) unitary transformations do not change entropy, and c) the function $g$ is an increasing function of its argument.

This is the case of a squeezed vacuum input \cite{Yuen}. The output of a phase independent amplifier with a squeezed vacuum input with squeezing parameter $r$, viz. with variance of the squeezed quadrature reduced by $\exp(-2r)$ from the vacuum state value, is equal to the output of an amplifier with gain $G'$ with vacuum state input followed by a squeezing operator with squeezing parameter $r'$ such that both $G \exp(-2 r)+(G-1) = \exp(-2 r') (2 G'-1)$ and $G \exp(2 r)+(G-1) = \exp(2 r') (2 G'-1)$ hold. The phase insensitive amplifier output with a squeezed vacuum input has the same entropy of the output with \textit{vacuum state} input of a phase insensitive amplifier with gain $(2G'-1)^2 = \left[G \exp(-2 r)+G-1\right] \left[G \exp(2 r)+G-1\right]$, or
\be G' = 1/2 + \sqrt{(G-1/2)^2 + G (G-1) \sinh^2(r)}. \ee
Therefore, the entropy of an amplifier output with a squeezed vacuum input is $g(G'-1)$ that, being $G' \ge G$, is $g(G'-1) \ge g(G-1)$. The same expression for the entropy is valid for a generic squeezed state, which can be obtained from the application of a unitary transformation (the displacement operator) to a squeezed vacuum with the same squeezing parameter. Being the average number of photons of the squeezed state \cite{Knight}
\be \langle a^\dagger a \rangle = \sinh^2(r) + |\langle a \rangle|^2, \ee
we may eliminate the squeezing parameter in the expression for $G'$ using $\sinh^2(r) = \langle a^\dagger a \rangle - |\langle a \rangle|^2$, obtaining
\be G' = \frac 1 2 + \left(G-\frac 1 2\right) \sqrt{1 + \frac{G (G-1)}{(G-1/2)^2} (\langle a^\dagger a \rangle - |\langle a \rangle|^2 )}. \label{260b} \ee
The above analysis confirms the earlier result \cite{Holevo} that the Holevo information (\ref{170}) (and the result can be extended to the case (\ref{180}) of an amplifier with spontaneous emission factor $n_\mathrm{sp} = N_\mathrm{ASE}/(G-1) > 1$) is the capacity if one restricts the class of input states to Gaussian states, including non classical ones.

\section{Infinitesimal gain case $G =1 + \epsilon$}

Let us now restrict the analysis to the case of infinitesimal gain $G = 1 + \epsilon$, and to a pure state input $|x \rangle \langle x|$. In this case, we may give an expression for the output entropy of the amplifier with any pure input state. Let us use for simplicity of analysis the expansion
\be \rho_0 = \int \df^2 \alpha P(\alpha) |\alpha\rangle \langle \alpha|, \label{PK} \ee
where $P(\alpha)$ is the Glauber-Sudarshan P-representation with trace-class norm of Klauder as in ref. \cite{Klauder}. Notice that since $P(\alpha)$ is a generalized distribution, Eq. (\ref{PK}) does not add any restriction to the class of input states.  The density matrix at the amplifier output is
\be \rho = \int \df^2 \alpha P(\alpha) \rho(\alpha), \ee
where $\rho=\rho(\alpha)$ is given by Eq. (\ref{55}).

To first order, if we set $G = 1 + \epsilon$, we may write
\bea \rho_\mathrm{ASE}(1+\epsilon) &=& \frac{1}{1+\epsilon} \sum_n \frac{\epsilon^n}{(1+\epsilon)^n} |n\rangle \langle n| \nonumber \\
&=& (1 -\epsilon) |0\rangle\langle0|+ \epsilon |1\rangle\langle 1| + O(\epsilon^2). \eea
Being
\be \rho = \int \df^2 \alpha  P(\alpha) D(\sqrt{1+\epsilon} \alpha) \rho_\mathrm{ASE}(1+\epsilon)  D^\dagger (\sqrt{1+\epsilon} \alpha), \ee
we obtain
\be \rho = (1-\epsilon) \rho_0  + \epsilon \rho_1 + O(\epsilon^2), \ee
where
\be \rho_0 = \int \df^2 \alpha  P(\alpha) D(\alpha) |0\rangle \langle 0|D^\dagger(\alpha) = \int \df^2 \alpha  P(\alpha) |\alpha \rangle \langle \alpha |,\ee
is the input state, which we assume pure $\rho_0^2 = \rho_0$, and
\bea \rho_1 &=& \int \df^2 \alpha  P(\alpha) \left\{ \epsilon^{-1} [D(\sqrt{1+\epsilon} \alpha)|0 \rangle \langle 0| D^\dagger(\sqrt{1+\epsilon} \alpha) \right. \nonumber \\
&& \left. - |\alpha\rangle \langle \alpha|] + D(\alpha) |1\rangle \langle 1| D^\dagger (\alpha)\right\}. \eea
Using that
\bea D(\sqrt{1+\epsilon} \alpha) &\simeq& D(\alpha) D(\epsilon \alpha/2) \nonumber \\
&\simeq&  D(\alpha)\left[ 1 + \frac \epsilon 2 (\alpha a^\dagger - \alpha^* a) \right], \eea
we obtain
\bea \rho_1 &=& \int \df^2 \alpha  P(\alpha) D(\alpha) \left[ \frac \alpha 2  |1 \rangle \langle 0 | \right. \nonumber \\
&& \left. + \frac {\alpha^*} 2  |0 \rangle \langle 1 | + |1\rangle \langle 1| \right] D^\dagger (\alpha), \eea
from which it is immediately apparent that
\be \mathrm{Tr}(\rho_1) = 1. \ee
Being
\be D(\alpha) a^\dagger = (a^\dagger - \alpha^*)D(\alpha), \ee
and $|1 \rangle = a^\dagger|0\rangle$, we have
\be \rho_1 = \rho_{1,1} + \rho_{1,2} +\rho_{1,3}, \ee
where
\bea \rho_{1,1} &=& \int \df^2 \alpha  P(\alpha) (a^\dagger - \alpha^*) |\alpha \rangle \langle \alpha | \frac \alpha 2,  \\
\rho_{1,2} &=& \int \df^2 \alpha  P(\alpha) \frac {\alpha^*} 2  |\alpha \rangle \langle \alpha | (a - \alpha), \\
\rho_{1,3} &=&  \int \df^2 \alpha  P(\alpha) (a^\dagger - \alpha^*) |\alpha\rangle \langle \alpha| (a - \alpha). \eea
The density matrices $\rho_0$ and $\rho_1$ are in general not orthogonal, they are so if and only if the input state is a coherent state. To show this result, let us compute Tr$(\rho_0 \rho_{1,1})$,
\bea \mathrm{Tr}(\rho_0 \rho_{1,2}) &=& \mathrm{Tr} \int \df^2 \alpha  P(\alpha) (a^\dagger - \alpha^*) |\alpha \rangle \langle \alpha | \frac \alpha 2 \rho_0 \nonumber \\
&=& \frac 1 2 \mathrm{Tr} \, \int \df^2 \alpha P(\alpha)  (a^\dagger - \alpha^*) a  |\alpha \rangle \langle \alpha | \rho_0  \nonumber \\
&=& \frac 1 2 \mathrm{Tr} \, a^\dagger a \int \df^2 \alpha  P(\alpha) |\alpha \rangle \langle \alpha | \rho_0 \nonumber \\
&& - \frac 1 2 \mathrm{Tr} \, a  \int \df^2 \alpha  P(\alpha) |\alpha \rangle \langle \alpha | a^\dagger \rho_0. \eea
We have then
\be \mathrm{Tr}(\rho_0 \rho_{1,2}) =  \frac 1 2 \left[ \mathrm{Tr} (a^\dagger a \rho_0^2) - \mathrm{Tr} (a \rho_0 a^\dagger \rho_0) \right]. \ee
For a pure state $\rho_0 = |\phi_0 \rangle \langle \phi_0|$, we have $\mathrm{Tr} (a^\dagger a \rho_0^2) = \mathrm{Tr} (a^\dagger a \rho_0) = \langle a^\dagger a \rangle$ and
\bea && \mathrm{Tr} (a \rho_0 a^\dagger \rho_0) =  \mathrm{Tr}(a |\phi_0 \rangle \langle \phi_0| a^\dagger |\phi_0 \rangle \langle \phi_0|) \nonumber \\
&=& \langle \phi_0| a |\phi_0 \rangle \langle \phi_0| a^\dagger |\phi_0 \rangle = |\langle a \rangle|^2, \eea
so that
\be \mathrm{Tr}(\rho_0 \rho_{1,2}) = \frac 1 2 \left[ \langle a^\dagger a \rangle - |\langle a \rangle|^2\right]. \ee
Using similar algebra for the other terms, we obtain $\mathrm{Tr}(\rho_0 \rho_{1,2}) = \mathrm{Tr}(\rho_0 \rho_{1,2})$ and $\mathrm{Tr}(\rho_0 \rho_{1,3}) = - 4 \mathrm{Tr}(\rho_0 \rho_{1,2})$, so that
\be \mathrm{Tr}(\rho_0 \rho_1) = - \left( \langle a^\dagger a \rangle - |\langle a \rangle|^2 \right). \ee
Being $ \rho = (1-\epsilon) \rho_0  + \epsilon \rho_1$ and $\mathrm{Tr}(\rho_0 \rho_1) = - \left( \langle a^\dagger a \rangle - |\langle a \rangle|^2 \right)$, we may decompose $\rho_1$ into a part parallel and another orthogonal to $\rho_0$. Using Gram-Schmidt orthogonalization, we may then complete the space with states orthogonal to both, whose contribution to $\rho$ is of second order in $\epsilon$ or higher. We may therefore restrict, to first order, the space of the amplifier output to two dimensions. The states that diagonalize $\rho$ to first order will therefore be a state parallel and another, $|\phi_1\rangle \langle \phi_1|$, orthogonal to $\rho_0 = |\phi_0\rangle \langle \phi_0|$, so that
\be \rho = (1-\epsilon')|\phi_0\rangle \langle \phi_0| + \epsilon' |\phi_1\rangle \langle \phi_1|, \ee
where $\langle \phi_0| \phi_1 \rangle = 0$ and $\epsilon'$ is proportional to $\epsilon$. Let us write $\rho_1$ in terms of $\rho_0 = |\phi_0\rangle \langle \phi_0|$ and $|\phi_1\rangle \langle \phi_1|$ as
\bea \rho_1 &=& a_0 |\phi_0\rangle \langle \phi_0| + a_1 |\phi_1\rangle \langle \phi_1| \nonumber \\
&& + b |\phi_1\rangle \langle \phi_0| + b^* |\phi_0\rangle \langle \phi_1|. \eea
We have $a_0 + a_1 = 1 \Rightarrow a_1 = 1 - a_0$, and $a_0 = \mathrm{Tr}(\rho_0 \rho_1) = - \left( \langle a^\dagger a \rangle - |\langle a \rangle|^2 \right)$, so that 
\bea \rho &=& \left[(1- \epsilon (1 - a_0) \right] |\phi_0\rangle \langle \phi_0| + \epsilon  (1 - a_0) |\phi_1\rangle \langle \phi_1| \nonumber \\
&& + \epsilon \left(b |\phi_1\rangle \langle \phi_0| + b^* |\phi_0\rangle \langle \phi_1| \right). \eea
After writing the secular equation, one notices that the off-diagonal terms give a contribution to the eigenvalues of the order $\epsilon^2$, so that  two eigenvalues of $\rho$ to first order in $\epsilon$ are $\lambda_1 = 1-\epsilon'$ and $\lambda_2 = \epsilon'$ with
\be \epsilon' = [1 + (\langle a^\dagger a \rangle - |\langle a \rangle|^2)] \epsilon. \ee
Using the diagonal representation, the entropy of the amplifier output is $S(\rho) = -\lambda_1 \ln \lambda_1 - \lambda_2 \ln \lambda_2$, hence
\be S(\rho) = -(1-\epsilon') \ln (1-\epsilon') - \epsilon' \ln \epsilon', \ee
which is minimal for states such that $\langle a^\dagger a \rangle = |\langle a \rangle|^2$, that is for coherent states inputs. In addition, any other state that is not a coherent state gives a larger output entropy.

An interesting check of this result may be obtained with Gaussian states, where we have shown that the entropy is given by $g(G'-1)$ where $G'$ is given by Eq. (\ref{260b}). If we set $G = 1 + \epsilon$, we have
\be G' \simeq 1 + \epsilon', \ee
where
\be \epsilon' = [1 + (\langle a^\dagger a \rangle - |\langle a \rangle|^2 )] \epsilon, \ee
hence the entropy of the amplified field is for small $G-1=\epsilon$
\be S(\rho) \simeq - \epsilon' \ln \epsilon', \ee
consistent with the general result.

A similar result can be obtained for an amplifier of arbitrary gain $G$ followed by large loss such that the overall gain of the cascade is $\epsilon (G-1)$, with $\epsilon \ll 1$. The density matrix of the attenuated field is
\bea \rho' &=& \int \df^2 \alpha  P(\alpha) D(\sqrt{\epsilon G} \alpha) \nonumber \\
&& \rho_\mathrm{ASE}[\epsilon(G-1) + 1]  D^\dagger (\sqrt{\epsilon G} \alpha). \eea
To first order in $\epsilon$ we have
\bea D(\sqrt{\epsilon G} \alpha) &=& 1 + \sqrt{\epsilon G}(\alpha a^\dagger - \alpha^* a) + \frac {\epsilon G} 2 (\alpha a^\dagger - \alpha^* a)^2 \nonumber \\
&=& 1 + \sqrt{\epsilon G}(\alpha a^\dagger - \alpha^* a) + \frac {\epsilon G} 2 [\alpha^2 a^{\dagger 2} + \alpha^{* 2} a^2 \nonumber \\
&&  - |\alpha|^2 (2 a^\dagger a + 1)], \eea
and
\be \rho_\mathrm{ASE}[\epsilon(G-1) + 1] = [1 - \epsilon(G-1)] |0\rangle \langle 0| + \epsilon(G-1) |1\rangle \langle 1|, \ee
We have
\bea \rho' &=& [1 - \epsilon(G-1)] [ |0\rangle \langle 0| \nonumber \\
&& + \sqrt{\epsilon G}  \int \df \alpha^2 P(\alpha) \left(\alpha |1\rangle \langle 0| + \alpha^* |0\rangle \langle 1| \right) \nonumber \\
&& + \int \df \alpha^2 P(\alpha) \epsilon G |\alpha|^2 |1\rangle \langle 1| \nonumber \\
&& + \epsilon G \int \df \alpha^2 P(\alpha) (\alpha^2 \sqrt 2 |2 \rangle \langle 0| + \alpha^{* 2} \sqrt 2 |0 \rangle \langle 2|  \nonumber \\
&& - |\alpha|^2 |0\rangle \langle 0|) + \epsilon(G-1) |1\rangle \langle 1|, \eea
that is
\bea \rho' &=& [1 - \epsilon(G-1)] [ |0\rangle \langle 0| \nonumber \\
&& + \sqrt{\epsilon G}  \left(\langle a \rangle |1\rangle \langle 0| + \langle a^\dagger \rangle |0\rangle \langle 1| \right) \nonumber \\
&& + \epsilon G \langle a^\dagger a \rangle |1\rangle \langle 1|  \nonumber \\
&& + \epsilon G (\langle a^2 \rangle \sqrt{2} |2 \rangle \langle 0| + \langle a^{\dagger 2} \rangle \sqrt{2} |0 \rangle \langle 2| - \langle a^\dagger a \rangle |0\rangle \langle 0|) \nonumber \\
&& + \epsilon(G-1) |1\rangle \langle 1|, \eea
The eigenvalues to first order in $\epsilon$ are solution of the secular equation
\bea && \lambda^2 - (1 - \epsilon G \langle a^\dagger a \rangle) \lambda + \epsilon [1 - \epsilon(G-1) - \epsilon G \langle a^\dagger a \rangle]  \nonumber \\
&&[(G-1) + G \langle a^\dagger a \rangle] -  \epsilon G |\langle a \rangle|^2 = 0 \eea
The zeroth and first order solution in $\epsilon$ are $\lambda_0 = 1 + O(\epsilon)$ and
\be \lambda_1 = \epsilon [(G-1) + G (\langle a^\dagger a \rangle -  |\langle a \rangle|^2)] + O (\epsilon^2). \ee
The entropy of the amplified and attenuated field is
\be S(\rho') \simeq - \lambda_1 \ln \lambda_1, \ee
which is minimum for $\langle a^\dagger a \rangle =  |\langle a \rangle|^2$, that is for a vacuum or coherent state input. Notice the case $G = 1$, that is, the case of no amplifier. In this case, the output entropy is zero for a coherent state, and non-zero for a different input state. This corresponds to the known result that an attenuated coherent state is still a coherent state, hence a pure state, whereas a non coherent state does not preserve its purity when attenuated.

\section{Conclusions}

We gave an alternative proof of the expression of the quantum equivalent of the capacity of a Gaussian channel with noise when we restrict the class of input states to classical input states and to Gaussian states.  We also \textit{directly} showed, without restricting the class of input states, that this expression holds true for a generic input state but for a phase insensitive amplifier of infinitesimal gain and for a phase insensitive amplifier of arbitrary gain after large attenuation. The first of these results shows that if the property is uniform with $G$, then there is an interval from $G = 1$ where the expression for the capacity is valid in general.


\begin{thebibliography}{99}

\bibitem{Giovannetti} V. Giovannetti, S. Guha, S. Lloyd, L. Maccone, J. H. Shapiro, and H. P. Yuen, ``Classical capacity of the lossy bosonic channel: the exact solution,'' Phys. Rev. Lett. \textbf{92}, 027902, 16 January 2004.

\bibitem{Shannon} C. E. Shannon, ``A mathematical theory of communication,'' The Bell System Technical Journal  \textbf{28}, 379--423 and 623--656, 1948.

\bibitem{Holevo} A. S. Holevo, M. Sohma, and O. Hirota, ``Capacity of Quantum Gaussian channels,'' \pra \textbf{59}, 1820--1828, 1999.

\bibitem{Lloyd} S. Lloyd, V. Giovannetti, L. Maccone, N.J. Cerf, S. Guha, R. Garcia-Patron, S. Mitter, S. Pirandola, M.B. Ruskai, J.H. Shapiro, and H. Yuan, ``The bosonic minimum output entropy conjecture and Lagrangian minimization,'' arXiv:0906.2758v3.

\bibitem{Mecozzi} A. Mecozzi, ``Quantum and semiclassical theory of noise in optical transmission lines employing in-line erbium amplifiers,'' J. Opt. Soc. Am. B \textbf{17}, 607--617, January 2000.

\bibitem{Gordon62} J. P. Gordon, ``Quantum Effects in Communications Systems,'' Proc. IRE 1898--1908 (1962).

\bibitem{Gordon64} J. P. Gordon, ``Noise at optical frequencies; information theory,'' in Proceedings of the International School of Physics ``Enrico Fermi,'' Course XXXI. London (UK): Academic Press, 1964, pp. 156--181.

\bibitem{Glauber} R. J. Glauber, ``Coherent and Incoherent States of the Radiation Field,'' Phys. Rev. \textbf{131}, 2766--2788, 1963.

\bibitem{Wehrl} A. Wehrl, ``General properties of entropy,'' Rev. Mod. Phys. \textbf{50}, 221--260, April 1978.

\bibitem{Yuen} H. P. Yuen, ``Two-photon coherent states of the radiation field,'' \pra \textbf{13}, 2226--2243, 1976.

\bibitem{Knight} P. L. Knight and V. Bu\v{z}ek, ``Squeezed states: Basic principles,'' in \textit{Quantum squeezing} P. D. Drummond and Z. Ficek (Eds.), Springer Series on Atomic, Optical, and Plasma Physics, Vol. 27, 2004.

\bibitem{Klauder} J. R. Klauder, ``Improved Version of Optical Equivalence Theorem,'' \textbf{16}, 2, 534--536, 21 March 1966.

\end{thebibliography}
\end{document}